\documentclass[10pt,b5paper,dvips]{article}
\usepackage{graphics}

\begin{document}
\usefont{OT1}{ptm}{m}{n}
\newpage
\null
\vskip 100pt
\begin{center}
 {\LARGE           GASDYNAMIC MODEL OF STREET CANYON             
                                \\[18pt]}
 {\large           Maciej M. Duras                     \\[6pt]}
 {\normalsize \it  Institute of Physics, 
                   Cracow University of Technology,                         
                   \\
                   ulica Podchorazych 1, PL-30-084 Cracow, 
                   Poland.                \\} 
		   Proceedings of the XIV Polish
		   Conference on Computer Methods in Mechanics (PCCMM'99), 
		   26th May 1999- 28th May 1999, 
		   Rzesz\'{o}w, Poland, 85-86 (1999).\\
		   AD 1999, July 26th
		   
\end{center}

\section{Abstract}
A general proecological urban road traffic control idea
for the street canyon is proposed
with emphasis placed on development of advanced
continuum field gasdynamical (hydrodynamical) 
control model of the street canyon.  
The continuum field model of optimal control of street canyon is studied.
The mathematical physics approach (Eulerian approach) 
to vehicular movement,
to pollutants' emission and to pollutants' dynamics is used.
The rigorous mathematical model is presented,
using gasdynamical (hydrodynamical) theory for both air constituents
and vehicles, including many types of vehicles and many types
of pollutant (exhaust gases) emitted from vehicles.
The six optimal control problems are formulated
and numerical simulations are performed.
Comparison with measurements are provided.
General traffic engineering conclusions are inferred.

\section{Description of the model}
The vehicular flow is multilane bidirectional one-level
one-dimensional rectilinear and it is considered
with two coordinated signalized junctions
\cite{Duras 1998 thesis, Duras 1996 Polish}.
The vehicles belong to different vehicular classes:
passenger cars, trucks.
The emission from vehicles are based on technical measurements
and many types of pollutants are considered
(carbon monoxide CO, hydrocarbons HC, nitrogen oxides NOx).
The vehicular dynamics is based on hydrodynamical approach
\cite{Michalopoulos 1984}.
The governing equations are continuity equation
for number of vehicles 
and Greenshields' equilibrium speed-density u-k model
\cite{Greenshields 1934}.

The model of dynamics of pollutants is also hydrodynamical.
The model consists of a set of mutually interconnected
vector nonlinear, spatially three-dimensional, temporally dependent, 
partial differential equations with nonzero righthand sides (sources),
and of boundary and initial problem.
The pollutants, oxygen, and remaining gaseous constituents of air,
are treated as mixture of noninteracting, Newtonian, viscous
fluid (perfect or ideal gases).
The model incorporates as variables the following fields:
density of mixture, mass concentrations of constituents of mixture,
velocity of mixture, temperature of mixture, pressure of mixture,
and intrinsic (internal) energy of mixture.
The energy of mixture is calculated
{\it ab initio} assuming classical Grand Canonical ensemble 
with external gravitational Newtonian body force density field of Earth.
The results from classical thermodynamics are adopted to
framework of thermodynamics of continuum, and finally 
used in hydrodynamical model.
The hydrodynamical model is based on assumption 
on local laws of balance (of conservation): 
of mass of mixture, of masses of constituents of mixture,
of momentum of mixture, and of energy of mixture.
The set of governing equations is complete 
and it is composed of the above balance equations,
of state equation (Clapeyron's law), and equations
of conservation of number of vehicles of given type.
The model of dynamics is solved by finite difference
scheme.

The six optimization problems are formulated
by defining functionals of total travel time,
of global emissions of pollutants, 
and of global concentrations of pollutants,
both in the studied street canyon,
and in its two nearest neighbour substitute canyons.
The vector of control is a five-tuple
composed of two cycle times, two green times,
and one offset time between the traffic lights.
The optimal control problem consists of
minimization of the six functionals
over admissible control domain.

The model depends on hydrodynamical parameters: 
velocity, density, temperature, pressure of mixture,
concentrations of constituents of mixture;
as well as on traffic parameters: jam, saturation vehicular densities,
free flow speed.
This variety of parameters produces variety of minimal (optimal) solutions.
The set of minimal (optimal) solutions for  a set of different hydrodynamical
and traffic parameters (scenarios) is presented, and discussion
of results is provided.
The comparison between numerical simulations and measurements
is presented.
Moreover, the inferences for traffic engineering are given.
The discussion of future development is also presented.

\end{document}